\documentclass[12pt]{article}
\usepackage{epsf}
\title{\bf Rotating QCD string and the meson spectrum}
\author{
V.L.Morgunov\thanks{morgunov@vxitep.itep.ru}, A.V.Nefediev
\thanks{nefediev@vxitep.itep.ru} and Yu.A.Simonov\thanks{simonov@vxitep.itep.ru}\\
Institute of Theoretical and Experimental Physics\\
117218, Moscow, B.Cheremushkinskaya
25, Russia}
\date{}
\newcommand{\be}{\begin{equation}}
\newcommand{\ee}{\end{equation}}
\begin{document}
\maketitle

\begin{abstract}

The spectra of light--light and heavy--light mesons are
described by spinless Salpeter equation and Dirac equation
respectively, which predict linear dependence of the meson mass squared
$M^2$ on angular momentum $J$ and number of radial nodes $n$. 
Both spectra are computed by the WKB
method  and shown to agree with exact numerical data within few
percent even for the lowest levels. The drawback of Salpeter and Dirac
equation is that (inverse) Regge slopes do not coincide with the string 
ones, $2\pi \sigma$ and $\pi\sigma$ respectively, because the string dynamics
is not taken into account properly.
The lacking string rotation is
introduced via effective Hamiltonian derived from QCD
which generates linear Regge trajectories for light
mesons with the correct string slope.
\end{abstract}

\section{Introduction}

QCD is believed to be the  fundamental  theory of strong
interactions and the meson spectroscopy is to be derived from QCD. The
spectrum of mesons has been treated in a sequence of models \cite{spec1,spec2}
which may be called QCD motivated, but still not directly derived
from the QCD Lagrangian.

The problem of the celebrated Regge behaviour of the hadron spectra has been
discussed in literature not once (see {\it e.g.} \cite{regge} and
references herein) but still attracts
considerable attention. 
The light--light meson spectrum obtained so far and reasonably describing
experiment can be written in the form
\be
M_{ll}^2(n,J)=(c_n^{(ll)}n+c_J^{(ll)}J+\Delta M^2_p+\Delta M^2_s),
\label{1}
\ee
where $n$ is the radial quantum number, $J$ being the total angular momentum,
$\Delta M^2_p$ contains the perimeter (self-energy) mass
correction as well as corrections to the first two terms, while $\Delta
M_s^2$ takes into account spin splittings.

For heavy--light mesons a similar relation holds true with the
subscript $ll$ changed for $hl$ in all coefficients:
\be
M^2_{hl}(n,J)=(c_n^{(hl)}n+c_J^{(hl)}J+\Delta M^2_p+ \Delta
M^2_s).
\label{2}
\ee

From general physical considerations one expects the spectra (\ref{1}) and (\ref{2}) to
follow from a string--like picture of confinement which predicts the
(inverse) Regge slope
\be
c_J^{(ll)}=2\pi\sigma,\quad c_J^{(hl)}= \pi \sigma.
\label{3}
\ee

Additional daughter Regge trajectories are given by
vibrational excitations missing in (\ref{1}) and (\ref{2}), which are due to hybrid
excitations, {\it i.e.} constituent
gluons attached to the fundamental string \cite{hybrid}. In what follows we are interested
only in radial and orbital excitations of the string.

The string slope (\ref{3}) is an important criterion and check for any
QCD inspired model since it requires the correct account for the rotation
of the string, which is not present in the potential models considered so
far. For example, relativistic spinless Salpeter equation with confinement
reduced to the linearly rising potential between quarks yields
\be
c_J^{(ll)}(Salpeter)=8\sigma,\quad c_J^{(hl)}(Salpeter)=4\sigma
\label{4}
\ee
that is about 25\% larger than (\ref{3}), whereas the one-body Dirac equation with the linearly
rising potential leads to
\be
c_J^{(hl)}(Dirac)=4\sigma,
\label{5}
\ee
if the potential is added to the energy term (vector confinement\footnote{Here we
leave aside the well-known problem of the Klein paradox revealing itself in case
of vector confinement.}), and
\be
c_J^{(hl)}(Dirac)=2\sigma,
\label{6}
\ee
for the potential added to the mass term (scalar confinement). Both results lead to
considerable discrepancies with (\ref{3}) and, as will be shown later,
this happens because the rotation of the string, and hence momentum dependence of the 
effective potential, is not taken into account.

It was found few years ago \cite{dks} that starting from the area law for
Wilson loops one arrives at the relativistic Hamiltonian for the
spinless quark and antiquark which possesses two different regimes: potential regime
for small angular momenta $L$ and any $n$, and
string--like one for large $L$ and fixed $n$. In the latter case the
dominant term in the Hamiltonian indeed describes the rotating QCD
string, so that the string Regge slope (\ref{3}) is readily reproduced.
Similar results were obtained independently  by numerical analysis
of the spinless quark--antiquark system \cite{numeric}.

In the present paper we concentrate on the quasiclassical approach to
mesons, as the WKB method allows to obtain analytic formulae for
the meson spectra of surprisingly high accuracy thus
giving evidence for the quasiclassical dynamics of confined quarks in the meson.

Therefore our first task will be to check the accuracy of
the WKB approximation for those cases where exact solutions are
feasible: spinless Salpeter equation for light--light mesons (potential
regime of the general QCD string formalism \cite{dks}) and the
Dirac equation for linear confining potential for the case of heavy--light
system. We argue that the accuracy of WKB results is very good even for
lowest states. However the slopes in both cases are incorrect, as in (\ref{4}) and (\ref{6})
respectively.

At this point we come to the main purpose of this study --- to include the proper
string dynamics, whereby abandoning the notion of local potential
and introducing a new entity, the QCD string, the effect which can not be
recasted in terms of local potential.
We use the Hamiltonian derived in \cite{dks} and calculate the quasiclassical spectrum
of light mesons. The results represent celebrated
straight-line Regge
trajectories even for low-lying states with the slope very close to the expected
string slope (\ref{3}).

In conclusion we demonstrate how other effects (spin and colour Coulomb interaction) can be included in the
same Hamiltonian to make a direct comparison with experiment.

\section{Meson spectrum and quasiclassical approximation}

We start with the spinless Salpeter equation which describes
relativistic quark and antiquark of equal masses $m$ with angular momentum $l=0$ and spin effects
neglected (see \cite{dks} for the derivation of this equation from the general meson
Green function in QCD).
\be
(2\sqrt{p_r^2+m^2}+\sigma r) \psi_n=M^{(ll)}_n\psi_n
\label{7}
\ee
The Bohr--Sommerfeld condition looks like
\be
\int^{r_+}_0 p_r(r) dr=\pi\left(n+\frac{3}{4}\right),\quad
n=0,1,2,\ldots,\quad r_+=\frac{M^{(ll)}_n-2m}{\sigma}
\label{8}
\ee
that yields
\be
M^{(ll)}_n\sqrt{\left(M^{(ll)}_n\right)^2-4m^2}-4m^2ln
\frac{\sqrt{(M_n^{(ll)})^2-4m^2}+M^{(ll)}_n}{2m}=4\sigma
\pi \left(n+\frac{3}{4}\right)
\label{9}
\ee

A similar consideration for the heavy--light system of masses $m$ and
$M$ ($M\to \infty$) gives
\be (\sqrt{p_r^2+m^2}+\sigma r)
\psi_n=M_n^{(hl)}\psi_n 
\label{10}
\ee
\be
M_n^{(hl)}\sqrt{\left(M^{(hl)}_n\right)^2-m^2}-m^2ln
\frac{\sqrt{\left(M^{(hl)}_n\right)^2-m^2}+M_n^{(hl)}}{m}=2\sigma \pi
\left(n+\frac{3}{4}\right).
\label{11}
\ee

Accuracy of WKB approximation (\ref{9}), (\ref{11}) can be tested $vs$ exact
solutions of the Salpeter equations (recently accuracy of WKB approximation was
checked for light--light mesons in \cite{wkb_new}). In Table 1 this
comparison is given for the light--light system with $m=0$ and heavy--light one with
$m_q=0.01~GeV$ and and $M_{\bar q}=10~GeV$. The mass $M_n^{(hl)}$ in the latter case
actually refers to the difference of the total mass of the heavy--light
system and the mass of the heavy antiquark.

\begin{table}[ht]
\hspace*{3cm}
\begin{tabular}{|c|c|c|c|c|c|c|}
\hline
$n$&0&1&2&3&4&5\\ 
\hline
$M_n^{(ll)}$(WKB)&1.373 & 2.097& 2.629& 3.070&3.455&3.802\\ 
\hline
$M_n^{(ll)}$(exact) & 1.412& 2.106& 2.634& 3.073& 3.457&3.803\\ 
\hline
$M_n^{(hl)}$(WKB)&0.971 & 1.483&1.859 &2.171 &2.443 & 2.688\\ 
\hline
$M_n^{(hl)}$(exact) &1.014 & 1.524& 1.917&2.246 &2.537&2.800\\ 
\hline
\end{tabular}
\caption{Comparison of the WKB and exact spectra for Salpeter equations 
(\ref{7}) for massless quark and antiquark and (\ref{10}) for  
the quark mass $m_q=0.01~GeV$ and antiquark mass $M_{\bar q}=10~GeV$.
In  both cases $\sigma=0.2~GeV^2$ and $l=0$.}
\end{table}

Summarizing, one can say that spectra (\ref{9}), (\ref{11}) (as function of $n$ for
$l=0$) indeed have the form (\ref{1}), (\ref{2}) with the corrections at large
$n$ in the form
\be
\Delta M^2=O\left(\frac{m^2}{M_n^2}ln\frac{M_n}{m}\right)=O\left(\frac{ln~n}{n}\right).
\label{12}
\ee

The WKB spectrum is linear in $n$ and its accuracy
is about 3-4\% even for the lowest state.

We now turn to the case of the Dirac equation with linear confining
potential studied in \cite{wkb1}.
The WKB method for the Dirac equation was
thoroughly investigated in \cite{wkb2} and recently applied to the
case of confining potential \cite{yas}. Let us briefly recall the results here.

The Dirac equation with scalar ($U$) and vector ($V$) local potentials has the form
\be
(\vec \alpha \vec p + \beta (m+U)+V)\psi_n= \varepsilon_n\psi_n,
\label{13}
\ee
and the WKB quantization condition is \cite{wkb2}
\be
\int^{r_+}_{r_-}\left(p+\frac{\kappa
w}{pr}\right)dr=\pi\left(n+\frac{1}{2}\right),\quad n=0,1,2,\ldots,
\label{14}
\ee
where
\be
p=\sqrt{(\varepsilon-V)^2-\frac{\kappa^2}{r^2}-(m+U)^2},
\label{15}
\ee
$$
w=-\frac{1}{2r}
-\frac{1}{2}\frac{U'-V'}{m+U+\varepsilon-V},
$$
$$
|\kappa|=j+\frac{1}{2}
$$

An approximate quasiclassical solution of (\ref{13}) obtained in \cite{yas} for the case 
$m=0$, $V=0$, $U=\sigma r$ is
\be
\label{16}
\varepsilon^2_n=2\sigma\left(2n+j+\frac{3}{2}+\frac{sgn\kappa}{2}
+\frac{\kappa\sigma}{\pi\varepsilon^2_n}\left(0.38+ln
\frac{\varepsilon^2_n}{\sigma|\kappa|}\right)+
 O\left(\left(\frac{\kappa\sigma}{\varepsilon_n^2}\right)^2\right)\right).
\ee

The last two terms on the r.h.s. of equation (\ref{16}) are sub-leading for large
$n$ and are generated by the term $\frac{\kappa w}{pr}$ (see (\ref{14})). 
One can see that the (inverse) Regge slope in $j$ in (\ref{16})
is equal to $2\sigma$ coinciding with the exact result (\ref{6}), but is not of string type.
As it was expected a $j$-independent scalar potential does not
describe the physical phenomenon of rotating string.

Still the accuracy of the WKB approximation is impressing. 
In Table 2 one can see the comparison of exact eigenvalues computed in
\cite{wkb1} with quasiclassical ones and with those obtained from (\ref{16}). 
The discrepancy is less then 1\% even for the lowest state and it is much 
better for higher states.

\begin{table}[ht]
\hspace*{3cm}
\begin{tabular}{|c|c|c|c|c|c|c|}
\hline
$n$ & $l$ & $j$ & $\kappa$ & $M_{\rm exact}$ & $M_{\rm WKB}$ & $M_{\rm eq.(\ref{16})}$ \\ 
\hline
0& 2& 3/2& 2& 1.209& 1.209& 1.208\\ 
\hline
0& 1& 1/2& 1& 1.024& 1.025& 1.025\\ 
\hline
0& 0& 1/2& -1& 0.725& 0.726& 0.733\\ 
\hline
0& 1& 3/2& -2& 0.957& 0.960&0.966\\ 
\hline
0& 2& 5/2& -3& 1.149& 1.149&1.155\\ 
\hline
\end{tabular}
\caption{Comparison of the exact spectrum for Dirac equation 
(\ref{13}) with WKB spectrum and approximate one given by equation (\ref{16}) 
for $m=0$, $V=0$ and $U=\sigma r$ with $\sigma=0.2~GeV^2$.}
\end{table}

\section{Rotating string in the spinless quark  Hamiltonian}

Let us turn back to the spinless Salpeter equation for the light--light meson
and take the non-zero angular momentum into account. 
As it was shown above the Salpeter equation with local $l$-independent 
potential leads to the incorrect Regge slope (\ref{4}), and therefore this case 
requires a special treatment. One needs a Hamiltonian taking into account dynamical
degrees of freedom of the string, {\it e.g.} in the form of time
derivatives of string coordinates. This was done explicitly in
\cite{dks}, where it was shown that starting from the QCD Lagrangian and writing 
the gauge invariant $q \bar q$
Green function 
for confined spinless quarks in the Feynman-Schwinger representation,
one can arrive at the Lagrange function of the system in the well-known form
\be
L(\tau)=-m_1\sqrt{\dot{x}_1^2}-m_2\sqrt{\dot{x}_2^2}
-\sigma\int_0^1d\beta\sqrt{(\dot{w}w')^2-\dot{w}^2w'^2},
\ee
where $\tau$ denotes the proper time of the system, 
the first two terms stand for quarks, whereas the last one describes the
minimal string with tension $\sigma$ developed between the constituents; 
$w_{\mu}(\tau,\beta)$ being the string coordinate. Adopting the straight-line anzatz
for the minimal string, {\it i.e.} $w_{\mu}(\tau,\beta)=\beta
x_{1\mu}+(1-\beta)x_{2\mu}$, synchronizing the quarks proper times,
$x_{10}=x_{20}=\tau=t_{\rm lab}$ and introducing auxiliary fields to get rid of the square
roots (see {\it e.g.} \cite{einbein}) one can obtain the following Hamiltonian in the centre of mass
frame (we consider the case of equal masses $m$)
$$
H=\frac{p^2_r+m^2}{\mu(\tau)}+
\mu(\tau)+\frac{\hat L^2/r^2}
{\mu+2\int^1_0(\beta-\frac{1}{2})^2\nu(\beta) d\beta}+
$$
\be
+\frac{\sigma^2
r^2}{2}\int^1_0\frac{d\beta}{\nu(\beta)}+
\int^1_0\frac{\nu(\beta)}{2}d\beta,
\label{17}
\ee
where the two auxiliary positive functions $\mu(\tau)$ and
$\nu(\beta,\tau)\equiv \nu(\beta)$ are to be varied and to be found from the
minimum of $H$ yielding quark energy and
string energy density respectively. A more detailed analysis of the role played by
auxiliary fields can be found in {\it e.g.} \cite{ysk}.

Note that Hamiltonian (\ref{17}) has a form of the sum of
\lq\lq kinetic" and \lq\lq potential" terms only due to auxiliary fields $\mu$ and
$\nu$. If one gets rid of them by substituting their extremal values, the
resulting Hamiltonian possesses a very complicated form which makes 
its analysis and quantization hardly possible. 

The centrifugal potential in Hamiltonian (\ref{17}) is of special interest to us 
and, most of all, the second term in the denominator. It is 
this term that describes extra inertia due to the string connecting the
quarks. Neglecting this term and taking extrema in the auxiliary fields one easily
arrives at the ordinary Salpeter Hamiltonian with linearly rising potential,
whereas account for this extra term describes the proper string rotation and brings the
slope of the Regge trajectory into correct form (\ref{3}). 
In the nonrelativistic expansion of Hamiltonian (\ref{17}) this term yields
the so-called string correction to the leading confining potential
$\sigma r$ \cite{dks}
$$
\Delta H_l=-\frac{\sigma \hat{L}^2}{6m^2r},
$$
the part of the interaction which explicitly depends on the angular momentum.

Hamiltonian (\ref{17})
assumes especially simple form in the case of zero angular momentum and 
after excluding the auxiliary fields produces Salpeter equation (\ref{7}).

Variation of (\ref{17}) over $\nu(\beta)$ gives the stationary
energy distribution along the string with $\beta\;(0\leq \beta \leq
1)$ being the coordinate along the string. Thus one obtains
\be
\nu_0(\beta)=\frac{\sigma r}{\sqrt{1-4y^2(\beta-\frac{1}{2})^2}},
\ee
where $y$ is to be found from the transcendental equation
\be
\frac{\hat L}{\sigma r^2}=\frac{1}{4y^2}
(arcsin~y-y\sqrt{1-y^2})+\frac{\mu{y}}{\sigma r},
\label{19}
\ee
and $\hat L^2=l(l+1)$.

Note that the maximal possible value of $y,\;
y=1$, yields the energy distribution $\nu_0^{free}(\beta)$ corresponding to the 
free open string (string without quarks at the ends) \cite{dks,numeric}.

In the general case inserting the extremal function $\nu_0(\beta)$ one obtains from (\ref{17})
\be
H=\frac{p^2_r+m^2}{\mu(\tau)}+\mu(\tau)+\frac{\sigma r}{y} arcsin~y+\mu (\tau)y^2
\label{20}
\ee
with $y=y(\hat L,r,\mu)$ defined by equation (\ref{19}).
Unfortunately no rigorous analytic calculations are possible anymore, so one has to
rely upon numerical calculations. But let us first perform some analysis of Hamiltonian
(\ref{20}).    

Neglecting $\mu$ in (\ref{19}) and $\mu y^2$ in (\ref{20}) (which is justified
for large $\hat L$ and $\sigma r$, so that $\frac{\mu}{\sigma r}\ll
1$) and varying over $\mu$ in (\ref{20}) one obtains
\be
H_{as}=2\sqrt{p^2_r+m^2}+\frac{\sigma r}{y} arcsin~y,
\ee
so that the second term on the r.h.s can be viewed as an effective potential, and we
would like to emphasize that this potential is non-trivially $l$-dependent.

In the general case one has a $\mu$-dependent Hamiltonian (\ref{17}) with the
\lq\lq potential" $U(\mu,r)$,
\be
U(\mu,r)=\frac{\sigma r}{y} arcsin~y+\mu y^2.
\ee

A simplifying approximation can be used at this step, namely
the standard WKB procedure can be applied to the Hamiltonian
\be
H=\frac{p^2_r+m^2}{\mu_0}+\mu_0+U(\mu_0,r),
\label{23}
\ee
which slightly differs from the exact Hamiltonian (\ref{20}) as
it treats $\mu_0$ as a variational parameter not depending on
$\tau$. We find eigenvalues $M(\mu_0,\hat L,n)$ and minimize them with
respect to $\mu_0$ to obtain the spectrum $M(\mu_0^*(\hat L,n),\hat L,n)$, where
$\mu_0^*(\hat L,n)$ being the extremal value of $\mu_0$.

To check the accuracy of such a procedure for the eigenvalues two Hamiltonians were considered:
\be
H_{1}=2\sqrt{p_r^2+m^2}+\sigma r,
\label{24}
\ee
\be
H_{2}=\frac{p_r^2+m^2}{\mu_0}+\mu_0+\sigma r,\quad\mu_0\;{\rm varied},
\label{25}
\ee
where $H_1$ is obtained from $H_2$ in the limit when $\mu_0\to\mu(\tau)$.  

The results are listed in Table 3. One can see that the accuracy of variational 
procedure (\ref{25}) is better than 5\% and it is reasonable even for $m$ tending 
to zero.

\begin{table}[ht]
\hspace*{3cm}
\begin{tabular}{|c|c|c|c|c|c|c|}
\hline
$n$&0&1&2&3&4&5\\ 
\hline
$M_n^{(1)}$& 1.475& 2.254& 2.825& 3.299& 3.713& 4.085\\
\hline
$M_n^{(2)}$& 1.412& 2.106& 2.634& 3.073& 3.457& 3.803\\
\hline
\end{tabular}
\caption{Comparison of the WKB spectrum of Hamiltonian (\ref{25}) $M^{(1)}_n$
with the exact spectrum of Hamiltonian (\ref{24}) $M^{(2)}_n$ for $m=0$ and $\sigma=0.2~GeV^2$.}
\end{table}

As a next step we use the standard WKB method to find the spectrum of
Hamiltonian (\ref{23}). To this end we write the 
Bohr--Sommerfeld condition as
\be
\int^{r_+}_{r_-} p_r (r)dr=\pi\left(n+\frac{1}{2}\right),
\label{27}
\ee
with
\be
p_r(r)=\sqrt{\mu_0(M-\mu_0-U(\mu_0,r))-m^2}.
\ee

The eigenvalues $M(\mu_0, \hat L, n)$ were found numerically from (\ref{27}) 
and the minimization procedure was then used
with respect to $\mu_0$. Results for $M_{nl}$ are given in Table 4
and depicted in Fig.1 demonstrating very nearly straight lines with 
approximately string slope $(2\pi\sigma)^{-1}$ in $l$ and as twice as smaller slope
in $n$.  

\begin{table}[ht]
\hspace*{3cm}
\begin{tabular}{|c|c|c|c|c|c|}
\hline
\hspace*{0.3cm}$n$  $l$&1&2&3&4&5\\ 
\hline
0&1.865&2.200&2.481&2.729&2.956\\ 
\hline
1&2.562&2.832&3.068&3.281&3.480\\ 
\hline
2&3.091&3.329&3.540&3.733&3.913\\ 
\hline
3&3.535&3.753&3.947&4.125&4.290\\ 
\hline
4&3.925&4.128&4.309&4.476&4.629\\ 
\hline
5&4.278&4.469&4.638&4.797&4.939\\ 
\hline
\end{tabular}
\caption{Quasiclassical spectrum of Hamiltonian (\ref{17}) for $m=0$ and
$\sigma=0.2~GeV^2$.}
\end{table}

\begin{figure}[ht]
\epsfxsize=12cm
\epsfbox{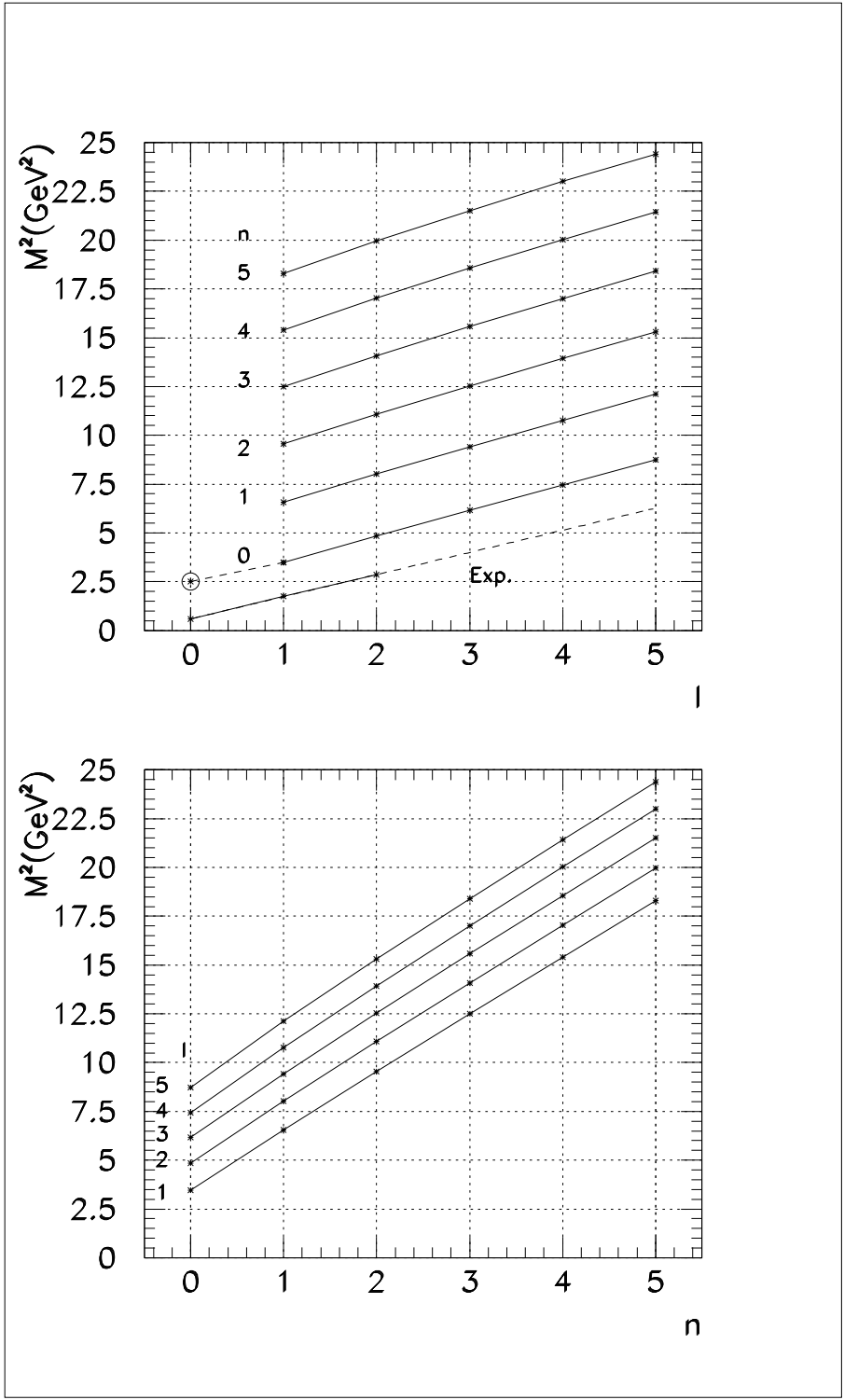}
\caption{Quasiclassical spectrum of Hamiltonian (\ref{17}) for $m=0$ and
$\sigma=0.2~GeV^2$. The leading experimental Regge trajectory in angular 
momentum $l$ is given in the upper plot for comparison. Note that as soon as $l=j-1$
for the given experimental trajectory, then it lies somewhat higher when
plotted in $l$ than when plotted in $j$. 
Theoretical prediction 
for the $\rho$-meson mass, $M^2_{\rho}\approx 2.5~GeV^2$, with colour Coulomb 
interaction and spin effects included, is shown not to violate the
straight-line behaviour of the leading theoretical trajectory.}
\end{figure}

\begin{figure}[ht]
\epsfxsize=13cm
\epsfbox{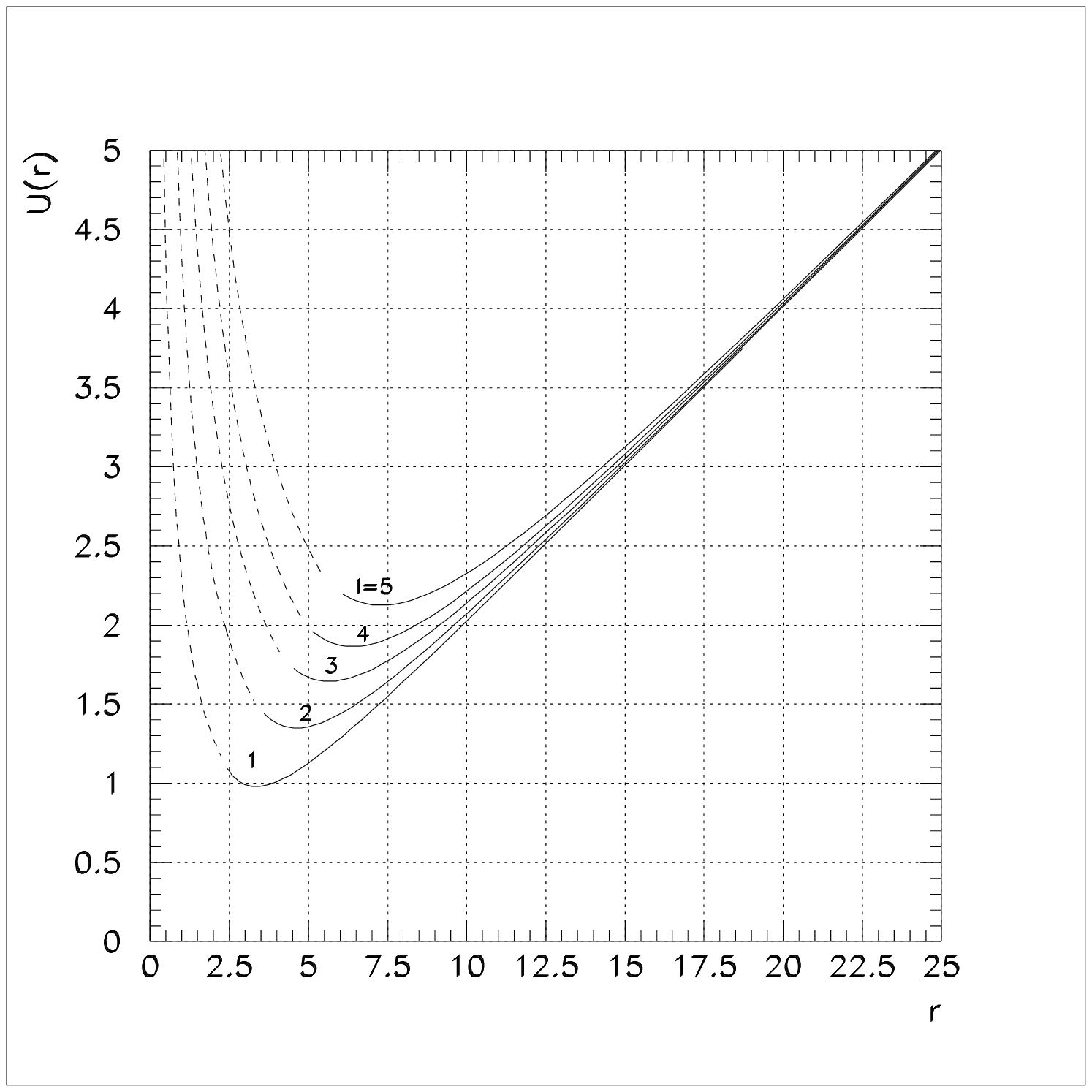}
\caption{Effective potential incorporating the string rotation as well as the quark radial motion for $\sigma=0.2~GeV^2$.
Dashed lines give matching with the centrifugal barrier at small $r$, whereas for all $l>0$ the 
effective potentials have the same 
asymptotics $\sigma r$ at large $r$. For $l=0$ the effective potential coincides with $\sigma r$ for all values of $r$.}
\end{figure}

Let us give a little comment concerning effective potential 
$U(\mu_0, r)$. Its behaviour at large and small distances can be extracted
analytically from Hamiltonian (\ref{17}) and coincides with that of the Salpeter: 
the centrifugal barrier at small $r$ and linear growth at large $r$.
Meanwhile in the region of intermediate values of $r$ this potential differs
from what one would have in the Salpeter equation and it is just this region which is
important to obtain the correct Regge slope. The form of the effective potential
is depicted in Fig.2 for several different angular momenta $l$. In 
case of $l=0$ the effective potential equals to $\sigma r$ for all values of $r$.

\section{Conclusion}

We have shown that the proper account of the string dynamics leads to practically linear
Regge trajectories, shown in Fig.1, with the slope numerically close to the conventional $(2\pi\sigma)^{-1}$.
The exact form of the effective potential incorporating the string rotation as well as the quark radial
motion was found numerically and shown in Fig.2.

To make contact with experimental data on meson masses one should specify corrections $\Delta M_p^2$, $\Delta
M^2_s$ in (\ref{1}), or in the case of the Hamiltonian formalism, one should add to Hamiltonian (\ref{20}) the
colour Coulomb term $V_C$ and spin-dependent interaction.

Treating the latter as perturbation one finds, {\it e.g.} for $\rho$ meson, a negative shift of the mass due to 
$V_C$ of about $160~MeV$ and positive correction of about $40~MeV$ due to hyperfine term $\Delta H_{ss}$.
Taking this into account one obtains for $\rho$ meson ($l=0$) the mass about $1.6~GeV$. It is clear
from Fig.1 that these corrections practically do not violate the linearity of Regge trajectories as $\rho$ meson 
lies on the continuation of the leading theoretical trajectory in $l$ (see dashed line attached to the trajectory with $n=0$). 
In this way starting from QCD and making one assumption of the area law for the Wilson loop we obtain linear Regge
trajectories for light quark mesons with the string slope. 

In this discussion quark spin effects have been taken into account perturbatively, which is a reasonable
approximation for the $\rho$ trajectory, but unacceptable for pions and kaons, since for the latter one needs
the full implementation of the chiral dynamics.

The progress in this direction was achieved in recent papers of one of the authors (Yu.S.) \cite{yas}, where
an effective Dirac equation for the quark moving in the field of an infinitely heavy antiquark source was
derived, and it was shown that solutions display the properties of confinement and chiral symmetry breaking.
The nonrelativistic limit of the resulting interaction lead to the conventional result for the confining term 
and to the spin-orbit interaction in agreement with the standard Eichten-Feinberg-Gromes results \cite{nonrel}.
Therefore pionic trajectories should be considered in this new formalism.

There is yet another question unanswered by our paper (and to our knowledge by all 
other existing papers) --- the
intercept of Regge trajectories $L_0\equiv L\;(M^2=0)$.
Theoretical intercept for the leading trajectory in $j$ (see Fig.1 and
the caption to it) is around -0.5, whereas it is +0.5 for the 
experimental $\rho$ trajectory also shown in Fig.1. The customary way
in the potential models is to add to the Hamiltonian a large negative constant 
$|C_0|\sim 1~GeV$ to reproduce
the intercept, but this would obviously violate the linearity of Regge trajectories.
Therefore one
expects that QCD provides a negative constant $\Delta M_p^2$ in (\ref{1}) but not 
in Hamiltonian (\ref{17}).

The authors are grateful to A.M.Badalian, A.B.Kaidalov, Yu.S.Kalashnikova and V.S.Po-pov for useful
discussions. Financial support of RFFI through the grants 97-02-16404, 97-02-17491 and 96-15-96740 is 
gratefully acknowledged.


\begin{thebibliography}{99}
\bibitem{spec1} D.R.Stanley and D.Robson, Phys.Lett.{\bf B45} (1980) 235\\
P.Cea, G.Nardulli and G.Preparata, Z.Phys. {\bf C16} (1982) 135,
Phys. Lett. {\bf B115} (1982) 310\\
J.Carlson, {\it et.al.} Phys.Rev. {\bf D27} (1983) 233\\
N.Isgur and S.Godfrey, Phys.Rev. {\bf D32} (1985) 189
\bibitem{spec2} J.L.Basdevant and S.Boukraa, Z.Phys. {\bf C28} (1983) 413
\bibitem{regge} G.F.Chew, Rev.Mod.Phys. {\bf 34} (1962) 394\\
I.Yu.Kobzarev, B.V. Martemyanov, M.G.Schepkin, UFN {\bf 162} (1992) 1 (in Russian)\\
M.G.Olsson Phys.Rev.{\bf D55} (1997) 5479\\
B.M.Barbashov and V.V.Nesterenko, {\it Relativistic string model
in hadron physics}, Energoatomizdat, Moscow, 1987 (in Russian)
\bibitem{hybrid} Yu.A.Simonov, in Proceedings of Hadron'93 (Como, 21-25 June 1993), ed. 
T.Bressani, A.Felicielo, G.Preparata and P.G.Ratcliffe;\\
Yu.S.Kalashnikova and Yu.B.Yufryakov, Phys. Lett. {\bf B359}
(1995) 175; Yad.Fiz. {\bf 60} (1997) 374
\bibitem{dks} A.Yu.Dubin, A.B.Kaidalov and Yu.A.Simonov, Phys.Lett. {\bf B323}
(1994) 41; Yad.Fiz. {\bf 56} (1993) 213
\bibitem{numeric} Dan La Course and M.G.Olsson, Phys.Rev. {\bf D39} (1989) 1751\\
C.Olsson and M.G.Olsson, MAP/PH/76/(1993)
\bibitem{wkb_new} V.D.Mur, B.M.Karnakov and V.S.Popov, in preparation, to be
submitted to ZhETP
\bibitem{wkb1} V.D.Mur, V.S.Popov, Yu.A.Simonov and V.P.Yurov, JETP {\bf 75} (1994) 1
\bibitem{wkb2} M.S.Marinov and V.S.Popov, JETP {\bf 67} (1974) 1250\\
V.L.Eletsky, V.D.Mur, P.S.Popov and D.N.Voskresensky, JETP {\bf 76}
(1979) 431; \\Phys.Lett. {\bf B80} (1978) 68
\bibitem{yas} Yu.A.Simonov, Yad. Fiz. {\bf 60} (1997) 2252
\bibitem{einbein}L.Brink, P.Di Vecchia, P.Howe, Nucl.Phys. {\bf B118} (1977) 76
\bibitem{ysk} Yu.S.Kalashnikova and A.V.Nefediev, Phys.At.Nucl. {\bf 60} (1997)
1389\\
Yu.S.Kalashnikova and A.V.Nefediev, Phys.At.Nucl. {\bf 61} (1998) 785
\bibitem{nonrel} N.Brambilla and A.Vairo, Phys.Lett. {\bf B407} (1997) 167\\
Yu.S.Kalashnikova and A.V.Nefediev, Phys.Lett. {\bf B414} (1997) 149
\end{thebibliography}
\end{document}